\documentclass[english,aps,prb,twocolumn]{revtex4}
\usepackage[T1]{fontenc}
\usepackage[latin1]{inputenc}
\usepackage{amsmath}
\usepackage{graphicx}
\usepackage{amssymb}

\makeatletter
\usepackage{babel}
\makeatother
\begin{document}

\title{Driven dissipative dynamics of spins in quantum dots}

\author{Frederico Brito$^{1}$\footnote{Permanent address: IBM T. J. Watson Research Center, NY US.}, Harry Westfahl Jr.$^{2}$, Amir O. Caldeira$^{1}$ and Gilberto Medeiros-Ribeiro$^{2}$ }

\affiliation{$^{1}$Instituto de F{\'\i}sica {}``Gleb Wataghin''{},
Universidade Estadual de Campinas, Caixa Postal 6165, 13083-970
Campinas, SP , Brazil,
\\
$^{2}$Laborat{\'o}rio Nacional de Luz S{\'\i}ncroton - ABTluS,
Caixa Postal 6192, 13043-090 Campinas, SP , Brazil }

\date{\today{}}

\begin{abstract}
We have studied the dissipative dynamics of a driven electronic
spin trapped in a quantum dot. We consider the dissipative
mechanism as due to the indirect coupling of the electronic spin
to acoustic phonons via the spin-orbit/electron-phonon couplings.
Using an effective spectral function of the dissipative phonon
bath, we evaluated the expectation values of the spin components
through the Bloch-Redfield theory. We show that due to a sharp
bath resonance present in the effective spectral function, with
typical energy much smaller than the electronic confinement
energy, the dissipative spin has a rich dynamical behavior that
helps us to determine some features of the spin-bath coupling. We
also quantify the effects produced by the sharp bath resonance,
and thus indicate the best regimes of operation in order to
achieve the longest relaxation times for the spin.
\end{abstract}

\pacs{Valid PACS appear here}

\maketitle

\section{Introduction}

During the last decade several physical systems have been proposed
as candidates for \emph{qubits} \cite{nielsen} and, among those, condensed matter devices have been attracting a growing
interest in the area \cite{guido}. There are many reasons for this
and most of them involve the way one designs, controls and accesses
individual qubits in quantum systems containing a very large
number of such entities. This is a natural consequence of the fact
that dealing with solid state devices one can use the same standard electronic circuitry of
conventional computers. Furthermore, new and powerful experimental
techniques allow us to expect that these qubits could be more
easily scalable and that control over the processing of
information and implementation of possible protocols of error
correction could be done in a more reliable and efficient way.
These are only a few points in favour of these systems as being good
candidates for qubits.

Despite these important favourable points, there are also negative aspects concerning the use of solid state
devices as qubits. Since they are mainly meso or nanoscopic devices,
which under specific conditions mimic two-state system behavior, it
is a very hard task to isolate them from their
environment. Moreover, as it is already well known, the system-bath
interaction causes loss of quantum coherence which is a drawback for
quantum computation \cite{CL,leggett}, in particular, if we do not
wish to operate our devices at too low temperatures. Therefore, there
must be a compromise between the desirable features possessed by these
systems and the undesirable effect of decoherence. It is along this
direction that people have been investigating electronic spins in
quantum dots as promising candidates for qubits \cite{lossdavid,gld}.
Since the confinement of the electronic wave function isolates the spins from most energy relaxation
channels, there are situations where
decoherence takes place only beyond a fairly long time interval. Here
we mean a time interval within which a large number of logical operations
could be coherently performed. In particular, for self-assembled quantum
dots, the energies involved in the spin dynamics are such that the
decoherence time is indeed very long. However, what is advantageous
on the one hand turns out to be a problem otherwise. As the spin is
a microscopic variable one has to face the problem of how to access it experimentally.

A possible way to address one of the many spins in the quantum processor
is by tuning the frequency of a time dependent external magnetic field
(the transverse pumping or control field) to the spin's Larmor frequency. Indeed, Koppens and collaborators \cite{Koppens}
have demonstrated very recently the first experimental realization of single
electron spin rotations in quantum dot systems, which is a necessary step for the implementation of
universal quantum operations. They showed, through current measurements in
double quantum dots, the ability of performing spin-flips using a resonant
oscillating magnetic field with the spin's Larmor frequency.  Therefore, in
this work, we shall investigate the dissipative dynamics of an electronic
spin confined to a quantum dot subject to a strong static magnetic
field and a much weaker transverse pumping field.

It has been demonstrated that the main mechanisms of relaxation of the latter
are the spin-orbit interaction \cite{alexander,ivchenko} and the
hyperfine interaction with the spins of the host lattice
\cite{erlingsson}. For static fields stronger than $1$T it is
expected that the spin-orbit interaction plays a major role in the
relaxation process. The relaxation times experimentally observed
for fields above $4$T are within the range
$\sim0.8-20$ms\cite{kroutvar,elzerman} which encourages us to
study the possibility of implementing spins confined in quantum
dots as qubits.

Following previous works \cite{harry}, we investigate the dissipative
effects originating from the coupling of the orbital electronic motion
to the acoustic phonon modes of the lattice where the dots are embedded.
Due to the spin-orbit coupling, the spin degree of freedom becomes
indirectly damped by the latter. Consequently, it is possible to define an effective dissipative
two-state system \cite{harry} whose dynamics may not be describable pertubatively \cite{khaetskii}.
In particular, one should stress the
existence of a very sharp effective bath resonance at energies much
lower than the planar electronic confinement energy.

The main goal of the present work is to evaluate the dissipative dynamics
of a structured spin-boson model in the presence of a time dependent
magnetic field aiming at the determination of the best operational
conditions under which these systems could be employed as good qubits.
In section \ref{model} we present the model we use to describe this
dynamical process. In section \ref{results}, we obtain the approximate analytical
solutions to the Bloch-Redfield equations for the average values of the spin components. Finally,
in section \ref{discussion}, we carry out a detailed analysis of the solutions
we have obtained.

\section{Model}
\label{model}

We consider quantum dots with strong confinement in the $z$
direction, and electronic orbital motion in the $x-y$ plane
subject to a confining parabolic potential. The spin-orbit
coupling is modelled by a Dresselhaus interaction term projected
onto the plane of the dot. Besides, we add a term due the presence
of an externally applied magnetic field $\bf B$. Thus, these
assumptions lead to the following spin-orbit Hamiltonian
\begin{eqnarray}
H_{SO} &=&  \hbar\omega_{0}\left(a_{x}^{\dagger}a_{x}+\frac{1}{2}\right)-
\beta\hat{\sigma}_{x} P_{x}\nonumber\\
& &+\hbar\omega_{0}\left(a_{y}^{\dagger}a_{y}+\frac{1}{2}\right)+\beta\hat{\sigma}_{y} P_{y}+
\frac{1}{2}g\mu_B{\bf{B}}\cdot{\boldsymbol{\hat{\sigma}}},\quad\label{Hso}
\end{eqnarray}
where $\omega_0$ is the lateral harmonic frequency, $\beta\equiv\gamma_{c}\langle k_{z}^{2}
\rangle=\gamma_{c}m^{\ast}\omega_{\bot}$
(with $\gamma_{c}$ is the Kane parameter\cite{ivchenko}, $m^{\ast}$
is the electron effective mass), $\hat{\boldsymbol{\sigma}}$ are the Pauli matrices, and $a_{x(y)}$
is the usual ladder operators for the $x(y)$ direction.

The electron-phonon Hamiltonian, considering either the
piezoelectric or the  deformation potential interactions with
acoustic phonon modes, can be mapped into the bath of oscillators
model \cite{CL} with the spectral function given by
\cite{harry}\begin{equation}
J_{s}\left(\omega\right)=m^{\ast}\omega_{D}^{2}\delta_{s}\left(\frac{\omega}
{\omega_{D}}\right)^{s}\theta\left(\omega_{D}-\omega\right),\label{Je-p}\end{equation}
 where $s=3$ for the piezoelectric interaction with dimensionless
coupling $\delta_{3}=\frac{(e_{m})_{14}^{2}\omega_{D}}{35\pi
m^{\ast}\rho}
\left(\frac{4}{3v_{t}^{5}}+\frac{1}{v_{l}^{5}}\right)$, and $s=5$
for the deformation potential with
$\delta_{5}=\frac{a_{c\Gamma}^{2} \omega_{D}^{3}}{2\pi\rho
m^{\ast}v_{l}^{7}}$, where $\omega_{D}$ is the Debye frequency.
$v_{l}$ and $v_{t}$ are the longitudinal and transverse sound
velocities respectively, $\rho$ is the material density,
$(e_{m})_{14}$ is the electromechanical tensor for zinc-blende
structures\cite{cardona}, and $a_{c,\Gamma}$ is the deformation
potential at the $\Gamma$ point\cite{cardona}. $\theta$ is the
Heaviside step function.

Provided that we have introduced a bath of oscillators for the electron-phonon
coupling, and because of the spin-orbit interaction, one can think
of this problem as a spin degree of freedom coupled to an effective
bath of oscillators \cite{harry} with the following effective bath spectral function
\begin{equation}
J_{eff}\left(\omega\right)=m^{\ast}\beta^{2}\frac{\delta_{s}\left(\frac{\omega}
{\omega_{D}}\right)^{s+2}}{Z\left(\omega\right)^{2}+\delta_{s}^{2}
\left(\frac{\omega}{\omega_{D}}\right)^{2s}}\theta
\left(\omega_{D}-\omega\right),\label{Jeff}\end{equation}
where
$Z\left(\omega\right)\equiv\left(\frac{\omega_{0}}{\omega_{D}}\right)^{2}-
\left(\frac{\omega}{\omega_{D}}\right)^{2}\left(1+\delta_{s}\phi_{s}\left(\frac{\omega}{\omega_{D}}\right)\right)$,
and
$\phi_{s}(x)\equiv\frac{2}{\pi}P\int_{0}^{1}x^{s}/(y^{3}-yx^{2})dy=-\frac{x^{s-2}}{\pi}(B(x,s,0)+(-1)^{s}B(-x,s,0))$,
with $B$ being the generalized \textit {incomplete beta function}.
\begin{figure}
\begin{center}\includegraphics[%
  width=0.9\columnwidth,
  keepaspectratio]{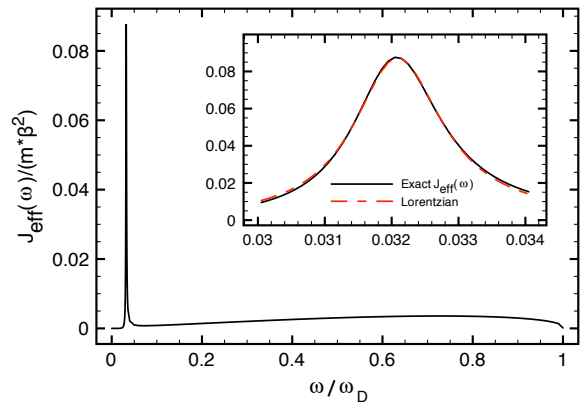}\end{center}
  \vspace{-0.6cm}
\caption{\label{j2} Spectral density $J_{eff}$ for piezoelectric
electron-phonon interaction, $s=3$. Inset: The region close to the
peak $\Omega_{s}$ can be well approximated by the Lorentzian in
Eq. \ref{Jlorentz} (dashed line). Here we assumed
$\frac{\omega_{0}}{\omega_{D}}\approx0.5$ and $\delta_{3}=355$.}
\end{figure}

Figure \ref{j2} presents the $J_{eff}(\omega)$ behavior for the
piezoelectric interaction case, $s=3$. The most prominent feature
of its behavior is the peak at $\omega=\Omega_s$, where $\Omega_s$
is determined by \begin{equation}
\Omega_{s}\approx\omega_{0}\sqrt{\frac{(s-2)}{(s-2)+\frac{2\delta_{s}}{\pi}}}.\label{Omegas}\end{equation}
Furthermore, the region around $\Omega_s$ can be well approximated
by a Lorentzian of width
$\ell\approx\frac{\omega_{D}\pi(s-2)}{4}\left(\frac{\omega_{0}}
{\omega_{D}}\sqrt{\frac{(s-2)\pi}{2\delta_{s}}}\right)^{s-1}$ (see
inset of Figure \ref{j2}), given by \begin{equation}
J_{eff}(\omega\approx\Omega_{s})\approx\frac{m^{\ast}\beta^{2}}{2}\left(\frac{\omega_{D}}
{\omega_{0}}\right)^{2}\frac{\Omega_s^{3}}{\omega_{D}}\left(\frac{\ell}{(\omega-\Omega_s)^{2}
+\ell^{2}}\right).\label{Jlorentz}\end{equation} The asymptotic
analysis of $J_{eff}(\omega)$ also reveals that in the low
frequency range, defined by $\omega\ll\Omega_s$ and
$\frac{\omega}{\omega_{D}}
\ll\left(\frac{\omega_{0}}{\omega_{D}}\frac{1}{\delta_s}\right)^{1/s}$,
the effective spectral density is always super-ohmic, with power
$s+2$. On the opposite side of the resonance, and in the high
frequency limit, $\Omega_s\ll\omega\ll\omega_{D}$, the spectral
density behaves as a power law in the frequency $\omega$ with
exponent $s-2$.

Since we know the spectral density of the effective bath of oscillators
coupled to the spin degree of freedom, our problem can now be modelled
by the driven spin-boson Hamiltonian
\begin{eqnarray}
H\left(t\right)&=&-\frac{\hbar}{2}[\Delta\hat{\sigma}_{x}+\epsilon\left(t\right)\,\hat{\sigma}_{z}]\nonumber\\
&&+\sum_{i}\hbar\omega_{i}\left(\hat{b}_{i}^{\dagger}\hat{b}_{i}+\frac{1}{2}\right)+\hat{\sigma}_{z}\sum_{i}c_{i}
\left(\hat{b}_{i}^{\dagger}+\hat{b}_{i}\right),\quad\label{Hsb}
\end{eqnarray}
where the spectral density of the environment is given by
(\ref{Jeff}), and the applied magnetic field ${\bf
B}=(-\hbar\Delta/g\mu_B,0,-\hbar\epsilon(t)/g\mu_B)$. In the weak
coupling limit, the equations of motion for the spin expectation
values, $\sigma_{i}\left(t\right)\equiv
Tr\left[\hat{\rho}\left(t\right)\hat{\sigma}_{i}\right]$, can be
written, for this specific model, as the generalized
Bloch-Redfield equation \cite{hartmann} 
\begin{eqnarray}
\dot{\sigma}_{x}\left(t\right)\!\!\!&=&\!\epsilon\left(t\right)\sigma_{y}-
\Gamma_{xx}\left(t\right)\sigma_{x}-\Gamma_{xz}\left(t\right)\sigma_{z}-A_{x}\left(t\right),\nonumber \\
\dot{\sigma}_{y}\left(t\right)\!\!\!&=&\!-\epsilon\left(t\right)\sigma_{x}+
\Delta\sigma_{z}-\Gamma_{yy}\left(t\right)\sigma_{y}-\Gamma_{yz}\left(t\right)\sigma_{z}-A_{y}\left(t\right),\nonumber\\
\dot{\sigma}_{z}\left(t\right)\!\!\!&=&\!-\Delta\sigma_{y},\label{BR}
\end{eqnarray}
where the fluctuating terms are given by \cite{hartmann}
\begin{eqnarray}
&&A_{x}\left(t\right)=\textrm{Im}F\left(t\right),~A_{y}\left(t\right)=\textrm{Re}F\left(t\right),~{\rm  with}\nonumber\\
&&F\left(t\right)=2\int_{0}^{t}dt'\textrm{Im}M(t-t')\,U_{RR}(t,t')U_{RL}(t,t'),\nonumber
\end{eqnarray}
and the temperature dependent relaxation rates are determined by\cite{hartmann}{\small
$$\Gamma_{ij}\left(t\right)=\int_{0}^{t}dt'\textrm{Re}[M(t-t')]b_{ij}(t,t'),~{\rm with}~ \Gamma_{yy}\left(t\right)=\Gamma_{xx}\left(t\right).$$}The correlation function 
$$M\left(t\right)=\frac{1}{\pi}\int_{0}^{\infty}d\omega
J(\omega)\frac{\cosh(\frac{\beta\omega}{2}-i\omega
t)}{\sinh(\frac{\beta\omega}{2})},$$ with $\beta\equiv1/k_{B}T$,
takes into account all the effects of the bath of oscillators. The
functions $b_{ij}$ read 
$$
b_{xz}=2\textrm{Re}\left[U_{RR}U_{RL}\right],~ b_{yz}=-2\textrm{Im}\left[U_{RR}U_{RL}\right]$$ and $b_{xx}=|U_{RR}|^{2}-|U_{RL}|^{2},$ where $U(t,t')$ is
the non-dissipative time evolution operator for the spin system
and its matrix elements $U_{RR}(t,t')=\langle R|U(t,t')|R\rangle$
and $U_{RL}(t,t')=\langle R|U(t,t')|L\rangle$,~are written in the
basis defined by
$\sigma_{z}|R\rangle\equiv+|R\rangle,\sigma_{z}|L\rangle\equiv-|L\rangle$.
Eqs. (\ref{BR}) are derived \cite{hartmann} under the
assumption that
$\frac{J(\omega)}{\omega}\ln(\Omega_{c}/\Delta)\ll1$, where
$\Omega_{c}$ represents the cutoff frequency of the system.

\section{Results}
\label{results}

From here onwards we will be interested in the study of a
monochromatic field of the form
$\epsilon\left(t\right)=2\epsilon_{0}\cos\Omega t$, considering it
as a small perturbation to the spin dynamics, \textit{i.e.},
$\epsilon_{0}/\Delta\ll1$. In order to solve the Bloch-Redfield
equations we need to compute the non-dissipative time evolution
operator $U(t,t')$. Using the rotating wave approximation (RWA),
we obtain the following simple analytic form to the time evolution
operator
\begin{equation}
U(t,t_{0})=R_{x}^{\dagger}\left(\Omega t\right)e^{\frac{i}{2}\epsilon_{1}(t-t_{0})
\hat{\boldsymbol{\sigma}}\cdot\mathbf{n}}R_{x}(\Omega t_{0}),\label{RWA}
\end{equation}
where $R_{x}\left(\Omega t\right)=e^{-\frac{i}{2}\Omega
t\hat{\sigma}_{x}}$,
$\hat{\boldsymbol{\sigma}}\cdot\mathbf{n}=\frac{\Delta-\Omega}{\epsilon_{1}}\hat{\sigma}_{x}+
\frac{\epsilon_{0}}{\epsilon_{1}}\hat{\sigma}_{z}=\cos\left(\phi\right)\hat{\sigma}_{z}+
\sin\left(\phi\right)\hat{\sigma}_{x}$, with
$\tan\left(\phi\right)\equiv\frac{\Delta-\Omega}{\epsilon_{0}}$
and
$\epsilon_{1}\equiv\sqrt{\epsilon_{0}^{2}+\left(\Delta-\Omega\right)^{2}}$.
$\epsilon_{1}$ and $\Delta$ are, respectively,  the Rabi and
Larmor frequencies of the problem, $\Omega$ is the external field
frequency and $\phi$ represents the frequency detuning. The
Bloch-Redfield coefficients can now be evaluated using the
spectral density of the bath oscillators and $U\left(t,t'\right)$.
We proceed evaluating the time integrals first. Those have sine
and co-sine elementary integrands that, when appropriately
rearranged, can be written in terms of $sinc$ functions, whose
maxima occur at the natural frequencies of the system $\Omega$, and
$\Omega\pm\epsilon_1$. Following this procedure, the coefficients
of the Bloch-Redfield equations can be written in the compact form
\begin{eqnarray}
A_{x}\left(t\right) & = & r_{1}\left(t\right)+m_{1}\left(t\right)
\sin\left(2\Omega t\right)+n_{1}\left(t\right)\cos\left(2\Omega t\right),\nonumber \\
A_{y}\left(t\right) & = & m_{2}\left(t\right)\sin\left(\Omega t\right)+
n_{2}\left(t\right)\cos\left(\Omega t\right),\nonumber \\
\Gamma_{xx}\left(t\right) & = & r_{3}\left(t\right)+m_{3}\left(t\right)
\sin\left(2\Omega t\right)+n_{3}\left(t\right)\cos\left(2\Omega t\right),\label{BRcomp}\\
\Gamma_{xz}\left(t\right) & = & m_{4}\left(t\right)\sin\left(\Omega t\right)
+n_{4}\left(t\right)\cos\left(\Omega t\right),\nonumber \\
\Gamma_{yz}\left(t\right) & = &
r_{5}\left(t\right)+m_{5}\left(t\right) \sin\left(2\Omega
t\right)+n_{5}\left(t\right)\cos\left(2\Omega t\right),\nonumber
\end{eqnarray} (see appendix for the explicit presentation of those
coefficients). The new coefficients $r_{i}\left(t\right),\,
m_{i}\left(t\right)$ and $n_{i}\left(t\right)$ have their time
dependence determined by the integrals $I_{n}$, Eqs.
(\ref{integrals1}) and (\ref{integrals2}). Those integrals, in the
regime of interest, $t\gg1/\omega_D$, have their main
contributions decomposed in two parts: one, time-dependent, occurs
due to terms arising from the poles of the spectral density in the
complex plane, which have lifetimes determined by their imaginary
parts; the other part is a time-independent contribution due to
the resonance of system with the external applied field.
Therefore, each coefficient $r_{i}\left(t\right),\,
m_{i}\left(t\right)$ and $n_{i}\left(t\right)$ can also be written
in well-characterized time dependent and independent parts. In
fact, considering $J(\omega)$ given by (\ref{Jeff}), whose  main
poles are $\Omega_s\pm i\ell$, we can write $r_1(t)$ as
\begin{eqnarray}
 r_{1}\left(t\right)  &\approx&  \ell J\left(\Omega_{s}\right)e^{-\ell t}\sum_{i=0}^{2}
 C_{i}\Bigg\{ \frac{\sin\left[\left(\Omega_{s}-\Omega^{(i)}\right)t\right]}{\Omega_{s}-\Omega^{(i)}} \nonumber\\
&&+\frac{\sin\left[\left(\Omega_{s}+\Omega^{(i)}\right)t\right]}{\Omega_{s}+\Omega^{(i)}}\Bigg\} +\tilde{r}_{1},\label{r1}
\end{eqnarray}
where
$\Omega^{(0)}\equiv\Omega,\,\,\Omega^{(1)}\equiv\Omega+\epsilon_{1}
\quad\text{and}\quad\Omega^{(2)}\equiv\Omega-\epsilon_{1}$;
$C_{0}\equiv\cos\left(\phi\right)^{2}/2,\,\, C_{1}\equiv\left(1+
\sin(\phi)\right)^{2}/4,\,\rm{and}~
C_{2}\equiv\left(1-\sin(\phi)\right)^{2}/4$. The first term on the
r.h.s of Eq. (\ref{r1}) is a direct contribution of the poles of
$J\left(\omega\right)$ , and the last term, $\tilde{r}_{1}$, takes
into account the time independent terms of $r_{1}\left(t\right)$.
Substituting (\ref{r1}) in the Bloch-Redfield equations, we can
find a particular solution for $\sigma_{x}$, up to first order in
$\ell$, due to the first term of Eq. (\ref{r1}). Thus we can write
$\sigma_{x}\left(t\right)$ as follows
\begin{eqnarray*}
\sigma_{x}\left(t\right) & \approx & \ell J\left(\Omega_{s}\right)e^{-\ell t}
\sum_{i=0}^{2}C_{i}\Bigg\{ \frac{\left(1-\cos\left(\Omega_{s}-\Omega^{(i)}
\right)t\right)}{\left(\Omega_{s}-\Omega^{(i)}\right)^{2}}\nonumber\\
&&+\frac{\left(1-\cos\left(\Omega_{s}+\Omega^{(i)}\right)t\right)}
{\left(\Omega_{s}+\Omega^{(i)}\right)^{2}}\Bigg\}
+\tilde{\sigma}_{x}\left(t\right).\end{eqnarray*} It is clear that
the particular solution has a lifetime determined by the width
$\ell$ of the peak of the effective spectral density. The modes
arising from this term oscillate with the system's natural
frequencies shifted by the bath resonance $\Omega_s$. Thus, the
contributions due to the poles of $J_{eff}(\omega)$ introduce a
new time scale, $1/\ell$, in the dynamics of the dissipative spin,
which distinguishes the short and long time regimes of the
problem. The first term of (\ref{r1}) is expected to be the
dominant contribution of the poles of $J_{eff}(\omega)$ for the
spin dynamics, because those arising from $m_1(t)$ and $n_1(t)$
oscillate with frequency $2\Omega$ larger than those from
$r_1(t)$. After integrating the equations of motion over any
measurable time interval, their effects become negligible.

Now, since we have determined the coefficients
$r(t),~m(t),~{\rm{and}}~n(t)$, we can obtain analytic solutions
for the Bloch-Redfield equations. For the long time regime,
$t>1/\ell$, those coefficients approach very quickly  their
asymptotic values $\tilde{f}\equiv f(t\rightarrow\infty)$, with
$f=r,~m,~{\rm{and}}~n$. A solution can be obtained using the
Laplace transformation for the regime of interest, namely
$\epsilon_{0}/\Delta\ll1$. Retaining terms up to order of
$\epsilon_{0}^{2}$, and considering the initial condition
$\sigma_{x}(0)=\pm1,\sigma_{y}(0)=0,\sigma_{z}(0)=0$, we find the
following approximated solutions
\begin{widetext}
\begin{eqnarray}
\sigma_{x}\left(t\right) & \approx & \ell J\left(\Omega_{s}\right)e^{-\ell t}
\sum_{i=0}^{2}C_{i}\left\{ \frac{\sin\left(\left(\Omega_{s}-
\Omega^{(i)}\right)t\right)}{\Omega_{s}-\Omega^{(i)}}+
\frac{\sin\left(\left(\Omega_{s}+\Omega^{(i)}\right)t\right)}
{\Omega_{s}+\Omega^{(i)}}\right\} +\sigma_{x}\left(\infty\right)+G\cos\left(2\Omega t\right)\nonumber \\
 &  & +\left(\sigma_{x}\left(0\right)-\sigma_{x}\left(\infty\right)-
 G-\sum_{i=0}^{2}H_{i}\right)e^{-\Gamma_{R}t}+\sum_{i=0}^{2}H_{i}
 e^{-\frac{\Gamma_{i}}{2}t}\cos\left(\omega_{i}t\right),\label{sigmax}\\
\sigma_{z}\left(t\right) & \approx & \frac{2\Delta}{\Omega}f
\left(\epsilon_{0},\Omega\right)\left(1-\cos\left(\Omega t\right)\right)
\left\{ \left(\sigma_{x}\left(0\right)-\sigma_{x}\left(\infty\right)-G-
\sum_{i=0}^{2}H_{i}\right)e^{-\Gamma_{R}t}+\sigma_{x}\left(\infty\right)\right\}\nonumber \\
 &  & +\Delta\sum_{i=0}^{2}H_{i}e^{-\frac{\Gamma_{i}}{2}t}
 \left\{ \frac{f\left(\epsilon_{0},\Omega-\omega_{i}\right)}
 {\Omega-\omega_{i}}\left(1-\cos\left(\left(\Omega-\omega_{i}\right)t\right)\right)+
 \frac{f\left(\epsilon_{0},\Omega+\omega_{i}\right)}{\Omega+\omega_{i}}
 \left(1-\cos\left(\left(\Omega+\omega_{i}\right)t\right)\right)\right\}.
\label{sigmaz}\end{eqnarray}
 \end{widetext}
where {\small
\begin{eqnarray}
\!\!\!&&\omega_{0}\approx2\Omega,\nonumber\\
\!\!\!&&\omega_{1}\approx\sqrt{\epsilon_{0}^{2}+\left(\Omega-\sqrt{\Delta\left(\Delta-2\gamma\right)}\right)^{2}}
\approx\sqrt{\epsilon_{0}^{2}+\left(\Omega-\Delta+\gamma)\right)^{2}},\nonumber\\
\!\!\!&&\omega_{2}
\approx\sqrt{\epsilon_{0}^{2}+\left(\Omega+\sqrt{\Delta\left(\Delta-2\gamma\right)}\right)^{2}}
\approx\sqrt{\epsilon_{0}^{2}+\left(\Omega+\Delta-\gamma\right)^{2}}.\nonumber
\end{eqnarray} }
The shift $\gamma$ in the natural resonance frequency can be
understood as the frequency correction to the spin dynamics due to
the interaction with the bath of oscillators. In contrast to the
procedure adopted in \cite{hartmann}, this correction arises
naturally from the solution obtained. The frequency shift $\gamma$
has the form of a Lamb shift
\begin{figure}
\begin{center}\includegraphics[%
  width=0.9\columnwidth]{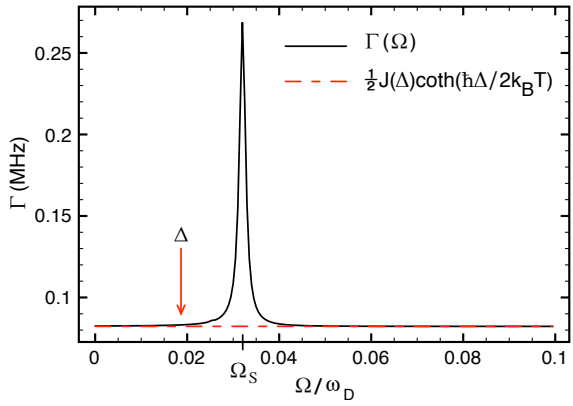}\end{center}
  \vspace{-0.6cm}
\caption{\label{gammaom} Rate $\Gamma$, Eq. \ref{gamma}, as a function
of the driven field frequency $\Omega$ (solid line). There is
a very pronounced peak at the region close to $\Omega_s$. Dashed
line represents the rate $\Gamma$ for the non-driven case, $\epsilon_{0}=0$.
Temperature used $T\approx1K$.}
\end{figure}
\begin{widetext}
\begin{equation}
\gamma\approx\tilde{r}_{5}=\sum_{i=0}^{2}\frac{C_{i}}{2\pi}\int_{0}^{\omega_{D}}d\omega\coth\left(\frac{\hbar\omega}
{2k_{B}T}\right)J\left(\omega\right)\text{P}\left[\frac{1}{\omega-\Omega^{(i)}}-\frac{1}{\omega+\Omega^{(i)}}\right].
\label{lambshift}\end{equation}
\end{widetext}
This expression is very similar to that found in the
non-perturbative treatment of the atom-field interaction
\cite{mandel}. However, it is worth noting that Eq.
(\ref{lambshift}) is derived here as a temperature dependent
function.

The rates appearing in Eq. (\ref{sigmax}) and (\ref{sigmaz}) can be written approximately as
$\Gamma_{R}\approx\left(1-\frac{2f^{2}(\epsilon_{0},\Omega)}{1+2f^{2}(\epsilon_{0},\Omega)}
\frac{\Delta(\Delta-2\gamma)}{\Omega^{2}}\right)\Gamma,$
 and  $\Gamma_{0}=2\Gamma_{1}=2\Gamma_{2}=2\Gamma$, where \begin{equation}
\Gamma\approx\tilde{r}_{3}=\frac{1}{2}\sum_{i=0}^{2}C_{i}J(\Omega^{(i)})
\coth\left(\frac{\hbar\Omega^{(i)}}{2k_{B}T}\right).\label{gamma}\end{equation}

Figure \ref{gammaom} presents the rate $\Gamma$ as a function of
the driven field frequency $\Omega$. It is clear that there is a
signature of the peak of spectral density
$J_{eff}\left(\omega\right)$ in its behavior. Close to the peak,
$\Omega\approx\Omega_s$, the rate $\Gamma$ can be several orders
of magnitude higher than the asymptotic value
$\frac{1}{2}J\left(\Delta\right)\coth\left(\hbar\Delta/2k_{B}T\right)$.
This result shows which regime of frequency must be avoided to
obtain the largest relaxation time of the system.

Finally, the coefficients can be written as 
\begin{eqnarray}
\sigma_{x}\left(\infty\right)&\approx &\left(\frac{1}{1+2f^{2}
(\epsilon_{0},\Omega)}\right)\tanh\left(\frac{\hbar\Delta}{2k_{B}T}\right),\nonumber\\
G&\approx&\frac{\epsilon_{0}}{\Omega}f\left(\epsilon_{0},\Omega\right)\sigma_{x}\left(\infty\right),\nonumber\\
H_{0}&\approx &-2H_{2}\approx-2G,~H_{1}\approx\left(\frac{\epsilon_{0}}{\omega_{1}}\right)^{2}\sigma_{x}\left(0\right).\nonumber
\end{eqnarray}
The function $$
f\left(\epsilon_{0},\Omega\right)=\epsilon_{0}\frac{\Omega}{\Omega^{2}-\Delta\left(\Delta-2\gamma\right)},$$
gives a measure of the effects of the driven field on the
spin dynamics. As expected, there are two important features to be
considered: one is its intensity $\hbar\epsilon_{0}/g\mu_{B}$; and
the other is how distant $\Omega$ is from the shifted natural
resonance frequency
$\sqrt{\Delta\left(\Delta-2\gamma\right)}\approx\Delta-\gamma$.

In the limit $\epsilon_{0}\rightarrow0$, we obtain
$\sigma_{x}\left(t\right)
\approx\sigma_{x}\left(\infty\right)+\left(\sigma_{x}\left(0\right)-\sigma_{x}\left(\infty\right)\right)e^{-\Gamma
t}$, where $\Gamma$ is the well-known expression\cite{leggett}
$\Gamma=\frac{1}{2}J(\Delta)\coth(\hbar\Delta/2k_{B}T)$,
and
$\sigma_{x}(\infty)=\tanh\left(\frac{\hbar\Delta}{2k_{B}T}\right)$
is exactly the expected thermodynamical mean value. For the
resonant case $\Omega\approx\Delta-\gamma$, we have
$f(\epsilon_{0},\Delta-\gamma)\rightarrow\infty$, which implies
$\sigma_{x}\left(t\right)\approx\sigma_{x}(0)e^{-\Gamma
t}\cos(\epsilon_{0}t)$. The system evolves with a frequency
imposed by the external driving field, and relaxes to a state
where both spin components are equally probable. A detailed
description of those phenomena is presented in the next section.

\section{\label{discussion}Discussion and conclusions}

\begin{figure}
\begin{center}\includegraphics[%
  width=0.9\columnwidth]{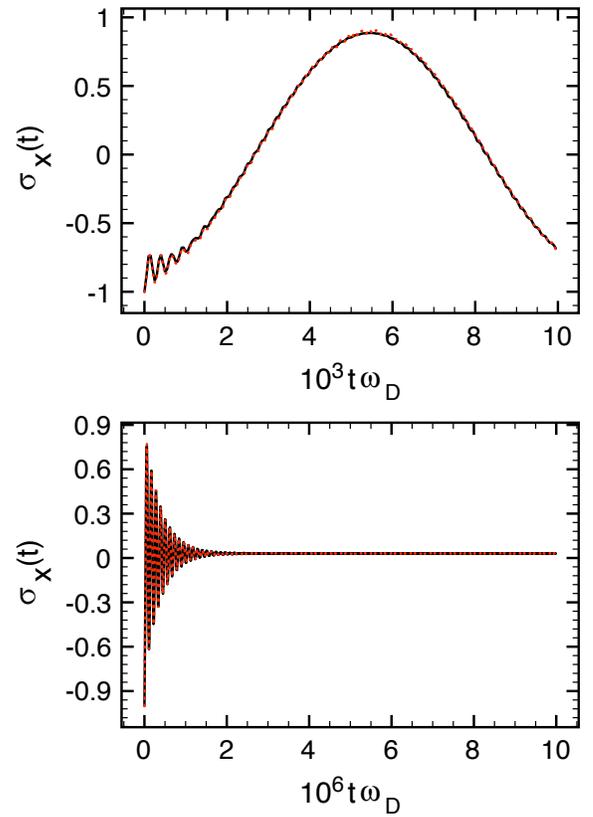}\end{center}
  \vspace{-0.6cm}
\caption{\label{irsigxplot} Time evolution of the expectation
value of $\hat{\sigma}_{x}$ for the near resonance case
$\Omega\approx\Delta-\gamma$. First plot presents the initial
dynamics, where  two well defined oscillation frequencies can be
seen. For times $t<1/\ell$, damped oscillations due the poles of
$J_{eff}(\omega)$, with relaxation time $1/\ell$, are observed. In
the long time regime, $t>1/\ell$, the system evolves with
 oscillation frequency $\epsilon_0$ and, initially, with large
amplitudes (reflecting the tuning between the frequency of the
driving field and the natural frequency of the system). Due to the
dissipative process, the system tends, in the time scale
$1/\Gamma$, to an equal probability state for spin up and down,
vanishing the average of $\hat{\sigma}_x$ (second plot). For very
long times, $t\gg1/\Gamma$, the system reaches a stationary state,
oscillating with frequency $2\Omega$ and very small amplitude.
Here we assumed $\omega_D/\ell=750$. Solid and dashed curves
represent the numeric and the analytic results obtained.}
\end{figure}

\begin{figure}
\begin{center}\includegraphics[%
  width=0.9\columnwidth]{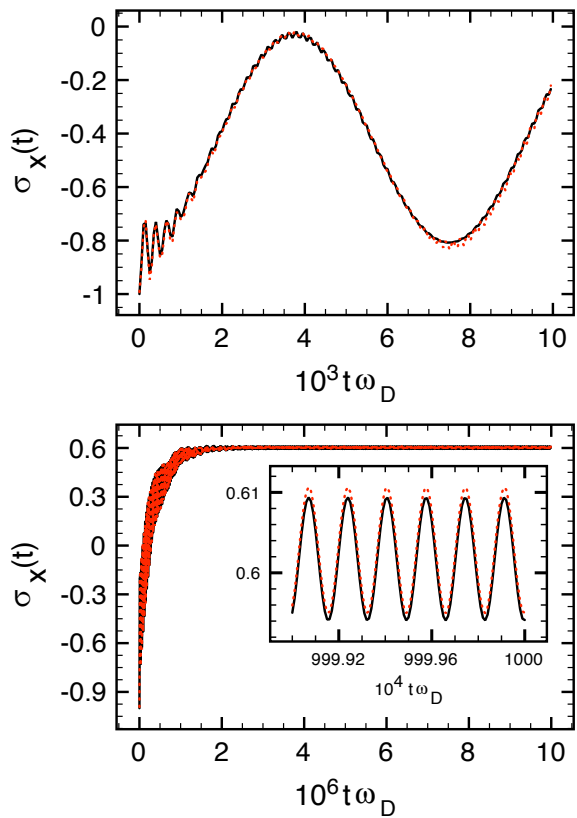}\end{center}
  \vspace{-0.6cm}
\caption{\label{irsigxforaplot} Time evolution of the expectation
value of $\hat{\sigma}_x$ for the  off-resonance $\Omega=\Delta$
case . Likewise the resonant case, the initial dynamics is
composed of two well defined oscillation frequencies (first plot).
Again, the contributions due the poles of $J_{eff}(\omega)$ vanish
after times $t\approx 1/\ell$. In the long time regime,
$t>1/\ell$, the system evolves oscillating with frequency
$\approx\sqrt{\epsilon_0^2+\gamma^2}$, and the amplitude observed
is not as larger as in the resonant case. The second plot presents
the dynamics for the long time regime. As one can see, the system
tends to approach the stationary state at the same time scale as
the resonance case, however we see that
$\sigma_x(t\rightarrow\infty)$ is now much closer to its
thermodynamical equilibrium value for $\epsilon_0=0$,
$\tanh(\hbar\Delta/2k_{B}T)\approx 1$. Once again, for very long
times, $t\gg1/\Gamma$, (inset second plot) the system oscillates
with frequency $2\Omega$ and very small amplitude. Here we assumed
$\omega_D/\ell=750$. Solid and dashed curves represent the numeric
and the analytic results obtained.}
\end{figure}

\begin{figure}
\begin{center}\includegraphics[%
  width=0.9\columnwidth,
  keepaspectratio]{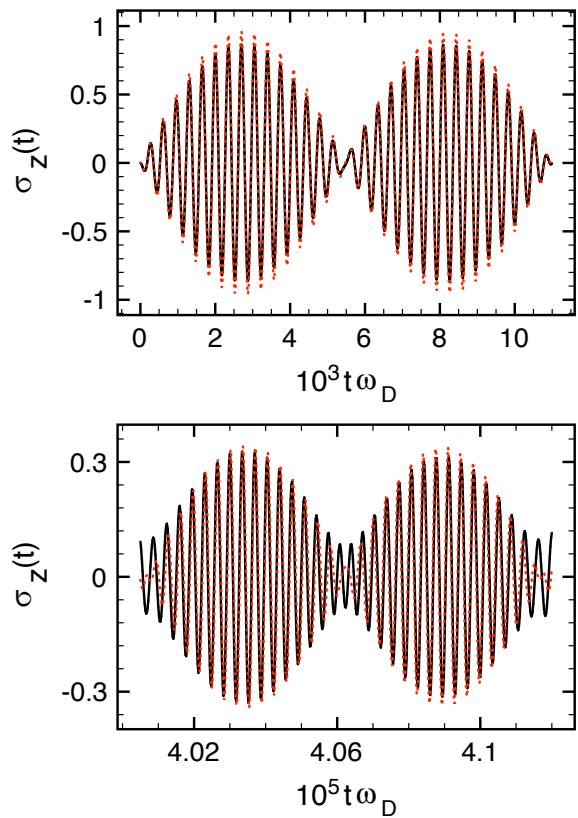}\end{center}
  \vspace{-0.6cm}
\caption{\label{irsigzresplot} Time evolution of the expectation
value of $\hat{\sigma}_{z}$ for the near resonance case
$\Omega\approx\Delta-\gamma$. The initial $\sigma_z$ dynamics
behavior (first plot) is characterized by a clear structure of
beats, with beat and angular frequency given by $\omega_1$ and
$\Omega+\omega_1/2$. Due the decoherence processes, the modes
related with the frequencies $\omega_i$ evanesce in the
characteristic time $1/\Gamma$, destroying the beat structure
(second plot), for which arises an oscillatory regime of same
frequency $\Omega$ of the applied field. Here we assumed
$\omega_D/\ell=750$. Solid and dashed curves represent the numeric
and the analytic results obtained.}
\end{figure}

\begin{figure}
\begin{center}\includegraphics[%
  width=0.9\columnwidth,
  keepaspectratio]{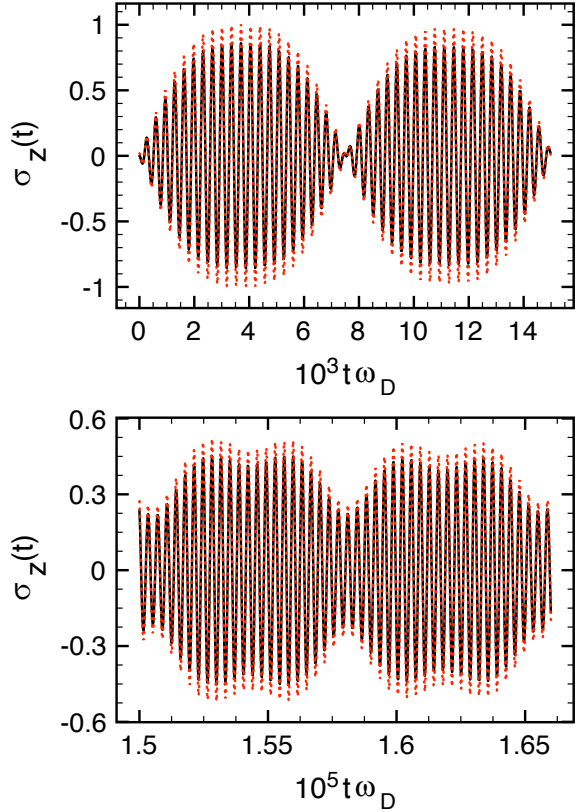}\end{center}
  \vspace{-0.6cm}
\caption{\label{irsigzforaplot} Time evolution of the expectation
value of $\hat{\sigma}_{z}$ for the off-resonance case
$\Omega\approx\Delta$. The same behavior for the resonant case is
observed. The initial $\sigma_z$ dynamics behavior (first plot) is
characterized by a clear structure of beats, with beat and angular
frequencies given by $\omega_1$ and $\Omega+\omega_1/2$. Due the
decoherence processes, the modes related with the frequencies
$\omega_i$ evanesce in the characteristic time $1/\Gamma$,
destroying the beat structure (second plot), for which  an
oscillatory regime of same frequency $\Omega$ of the applied field
arises. Here we assumed $\omega_D/\ell=750$. Solid and dashed
curves represent the numeric and the analytic results obtained.}
\end{figure}

Figures \ref{irsigxplot}-\ref{irsigzforaplot} present the
expectation  values of $\hat{\sigma}_{x}$ and $\hat{\sigma}_{z}$
for the near resonance, $\Omega\approx\Delta-\gamma$, and the
off-resonance, $\Omega=\Delta$, cases. Here we assume
$\hbar\Delta\gg k_{B}T$, $\Delta\approx\Omega_s/2$, $\epsilon_0/\Delta\approx0.03$ and
$\omega_{D}/\ell\approx750$.

\begin{figure}
\begin{center}\includegraphics[%
  width=0.9\columnwidth,
  keepaspectratio]{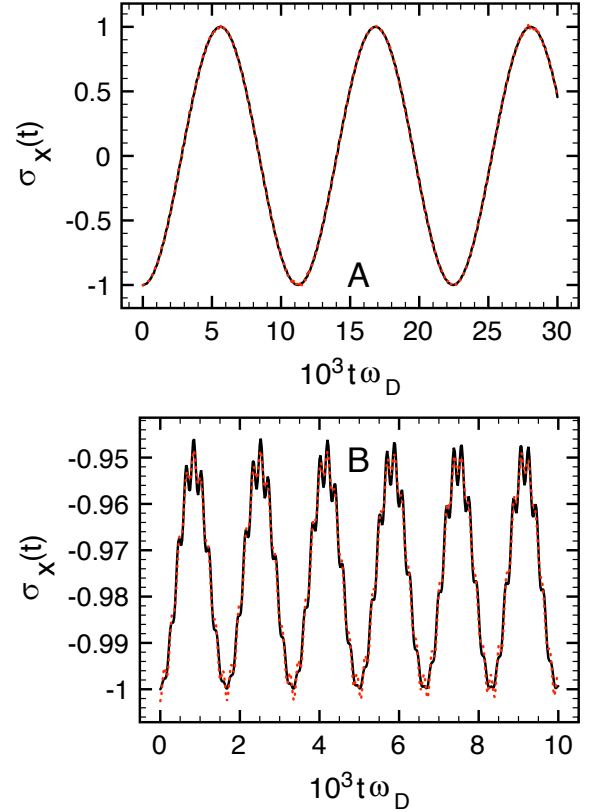}\end{center}
  \vspace{-0.6cm}
\caption{\label{sigxplot} The time evolution of the expectation
value $\sigma_{x}\left(t\right)$ using the approximate analytic
solution Eq.\ref{sigmax} (dashed line) and the exact numeric
calculation (solid line). Plot \textbf{A} presents the behavior
for the resonance case $\Omega\approx\Delta-\gamma$. The
oscillation frequency observed is $\omega_{1}\approx\epsilon_{0}$.
Plot \textbf{B} shows the case off resonance, $\Omega=0.8\Delta$.
The main oscillation frequency is $\omega_{1}$. The estimate
relaxation time for these plots is $T_{1}=12\mu s$. The
temperature used is $T\approx1K$. }
\end{figure}

\begin{figure}
\begin{center}\includegraphics[%
  width=0.9\columnwidth,
  keepaspectratio]{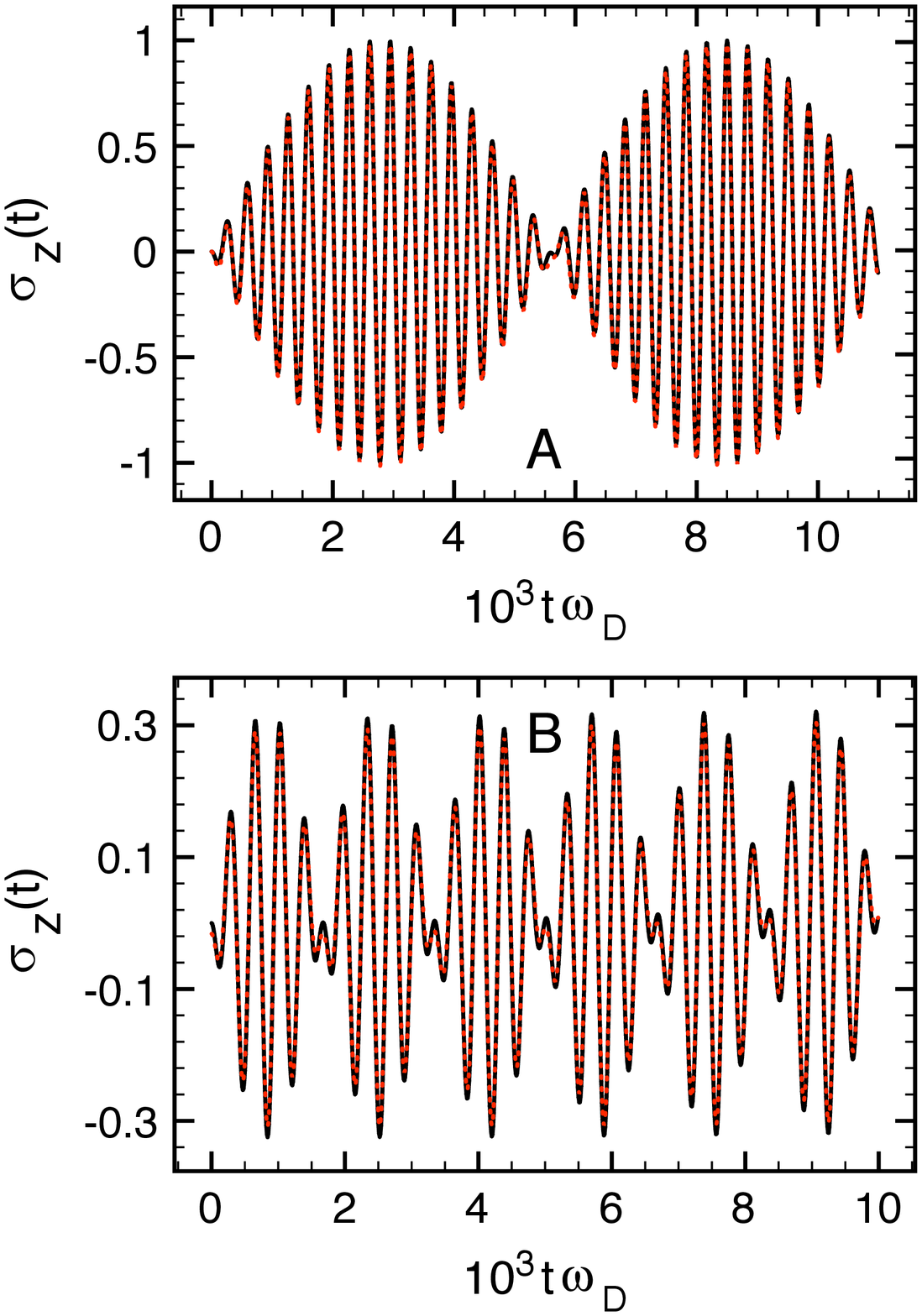}\end{center}
  \vspace{-0.6cm}
\caption{\label{sigzplot} The time evolution of the expectation
value $\sigma_{z}\left(t\right)$ using the approximate analytic
solution Eq.\ref{sigmax} (dashed line) and the exact numeric
calculation (solid line). Plot \textbf{A} presents the behavior
for the resonance case $\Omega\approx\Delta-\gamma$. $\omega_{1}$
and $\Omega-\omega_{1}/2$ are the beat and the angular frequencies
respectively. Plot \textbf{B} shows the case off resonance
$\Omega=0.8\Delta$. We can identify approximately the beat and
angular frequencies as $\omega_{1}$ and $\Omega+\omega_{1}/2$. The
temperature used is $T\approx1K$. }
\end{figure}

It is well known \cite{sakurai} that, in the absence of
dissipation, the system would indefinitely evolve into coherent
cycles of absorption and emission of energy - the Rabi
oscillations. Furthermore, the change in the spin up and down
populations exhibits an oscillatory time dependence with frequency
$\sqrt{\epsilon_0^2+(\Omega-\Delta)^2}$, and amplitude
$\epsilon_0^2/(\epsilon_0^2+(\Omega-\Delta)^2)$. It is worth
noting that under the resonance condition, $\Omega=\Delta$, the
oscillating field fully drives the transitions between up and down
spin states and that the weaker the driving field, the narrower
the resonance peak. As one can notice by inspecting figures
\ref{irsigxplot} and \ref{irsigxforaplot}, the coupling
system-bath changes some of those features. First of all, the
resonance condition is no longer verified at the frequency
associated with the static applied field. Now, the new natural
frequency is given by $\Omega=\Delta-\gamma$, where the shift
$\gamma$, Eq. (\ref{lambshift}), is completely associated with the
coupling between the system and its reservoir - Lamb shift. This
is exactly the behavior we have for the first plots of figures
\ref{irsigxplot} and \ref{irsigxforaplot}: the largest amplitudes
of oscillation occur for those frequencies closest to the new
resonance condition. As expected, the component $\sigma_x$
oscillates with frequency $\epsilon_0$ at  resonance, and
$\sqrt{\epsilon_0^2+\gamma^2}$ for the off-resonance
$\Omega=\Delta$ case. Another important change in the spin
dynamics is that the system does not evolve indefinitely in
coherent cycles of absorption and emission of energy. In fact, the
dissipation process induced by the coupling to the dissipative
bath destroys the coherence in the emission-absorption cycles and,
after a long time,  decoherence takes place. Indeed, the behavior
of the expectation value of $\sigma_x(t)$ in the long time regime
(figures \ref{irsigxplot} and \ref{irsigxforaplot}), reveals this
decoherence process. As one can see, the large amplitude of
oscillations observed initially, a consequence of the coherent
emission-absorption stimulated by the field, tends to decrease in
time. This process occurs in a time scale given by $1/\Gamma$, Eq.
(\ref{gamma}), which has almost the same value as presented in
figures \ref{irsigxplot} and \ref{irsigxforaplot}. This happens
because for both cases we are away from the peak of the rate
$\Gamma$, Figure \ref{gammaom}, where it is practically a constant
given by $\frac{1}{2}J(\Delta)\coth(\hbar\Delta/2k_BT)$. For very
long times, $t\gg1/\Gamma$, the system reaches a stationary state,
oscillating with frequency $2\Omega$ (inset of figure
\ref{irsigxforaplot}) and small amplitude around the asymptotic
value $\sigma_x(t\rightarrow\infty)$. Here we observe that this
asymptotic value changes dramatically when we leave the resonance
condition. This feature can be understood as follows. At
resonance, the driving field causes perfect transitions between up
and down eigenstates of $\hat{\sigma}_x$, but because the
system-bath coupling destroys the coherence along the process, we
see that after a very long time, $t\gg1/\Gamma$, a completely
random  emission-absorption process sets in and makes the
expectation  value of $\hat{\sigma}_x$ vanish. For the
off-resonance case, the driving field no longer produces 100$\%$
probability transitions in the system. Again, the coupling with
the bath destroys the coherence of this process, and favours the
occupation of the lowest energy state. Thus, as we are considering
a small applied field, i.e, $\epsilon_0/\Delta\ll 1$, the bath
drives the system towards its thermodynamical equilibrium state at
zero driving field,
$\sigma_x\rightarrow\tanh\left(\frac{\hbar\Delta}{2k_B T}\right)$.
Since we have assumed $\hbar\Delta\gg k_BT$, this implies
$\tanh\left(\frac{\hbar\Delta}{2k_BT}\right)\approx1$. How close
the system will approach that asymptotic value depends on both the
intensity and frequency of the driving field.

All the features discussed so far were general consequences of the
 system-bath interaction, and do not depend on the specific form of
the spectral density of the bath. Nevertheless, the first plots of
figures \ref{irsigxplot} and \ref{irsigxforaplot} show a
particular behavior for short times: two well defined regimes of
oscillation frequencies can be observed. This phenomenon is a
signature of the rich structure of the effective spectral density,
Eq. (\ref{Jeff}). In fact, because $J_{eff}(\omega)$ has poles in
the complex plane, new modes of oscillation arise in the spin
dynamics. Those new modes have lifetimes imposed by the imaginary
part of the poles, whereas the natural  frequencies of the system
are shifted by their real part. As we could observe from
$J_{eff}(\omega)$ (\ref{Jeff}), the most important modes ( those
with  long lifetimes and measurable amplitude of oscillations) due
to the poles are associated with the peak of $J_{eff}(\omega)$, at
which a strong correlation between orbital and spin degrees of
freedom takes place. These modes, (see the first term of Eq.
(\ref{sigmax})), have lifetimes determined by the width $\ell$ of
the bath resonance, and frequencies $\Omega_s-\Omega^{(i)}$, with
$\Omega^{(0)}=\Omega$, $\Omega^{(1)}=\Omega-\epsilon_{1}$ and
$\Omega^{(2)}=\Omega+\epsilon_{1}$. The amplitude of oscillation,
assuming a weak driving field, $\epsilon_0/\Delta\ll 1$, can be
approximated by $\frac{2\ell
J(\Omega_s)}{(\Omega_s-\Delta)^{2}}$ (for
$\Delta\neq\Omega_s$). Therefore, if one experimentally reaches
the conditions $\frac{2\ell
J(\Omega_s)}{(\Omega_s-\Delta)^{2}}\approx O(1)$ and
$\ell/\Delta\ll 1$, it would be possible to determine some
physical parameters of spin-bath coupling, {\textit{e.g.}}, the
electron-phonon constant coupling $\delta_s$.

Figures \ref{irsigzresplot} and \ref{irsigzforaplot} show the
$\sigma_{z}$ component for the same conditions previously
discussed. As it can be noticed, there is a clear structure of
beats in its dynamical behavior. At resonance, figure
\ref{irsigzresplot}, the beat and angular frequencies are,
respectively, given by $\omega_{1}$ and $\Omega+\omega_{1}/2$. For
this component we do not verify two oscillatory regimes for times
of the order of $1/\ell$. This occurs because of the suppression
of the pole's contribution by the strong oscillations of the
remaining terms. As a consequence of the decoherence process
imposed by the bath of oscillators, other important feature is
that, for long times, the modes related with the frequencies
$\omega_{i}$ evanesce, destroying the beat structure. These modes
are associated with the coherent oscillations due to the driving
field. Thus, after a long time, only an oscillatory regime of
frequency $\Omega$ remains. The characteristic time for that is of
the order of $1/\Gamma$. The regime slightly off  resonance
$\Omega=\Delta$, figure \ref{irsigzforaplot}, practically does not
change. In this case the beat and angular frequencies are given
respectively by $\omega_{1}$ e $\Omega-\omega_{1}/2$. This shows
that the component $\sigma_{z}$ is not so sensitive to how close
the external field frequency is to the natural resonance frequency
of the spin. As it is directly coupled to the external field, this
component has a stronger dependence on its amplitude.

Now, we will focus on the specific case in which we use bulk
physical parameters for typical quantum dots size. We assume here
the piezoelectric electron-phonon interaction , $s=3$, and quantum
dots frequencies $\omega_{0}\equiv15$meV.
The external fields applied to the dots are assumed to be such
that $\hbar\Delta/k_{B}=6.7K$ and $\hbar\epsilon_{0}/k_{B}=0.2K$,
and the Debye frequency $\hbar\omega_{D}/k_{B}=360K$. Figures
\ref{sigxplot} and \ref{sigzplot} present the $\sigma_{x}$ and
$\sigma_{z}$ dynamics using the bulk physical parameters
$\beta=3000m/s$, $m^{\ast}=0.063m_{e}$ and $\delta_{3}=355$. The
temperature used for the calculation is $T\approx1K$.

The plots \textbf{A} in figures \ref{sigxplot} and \ref{sigzplot}
show the dynamics of the expectation values
$\sigma_{x}\left(t\right)$ and $\sigma_{z}\left(t\right)$,
respectively, for the case close to the resonance,
$\Omega\approx\Delta-\gamma$. We can see that the dynamics of
$\sigma_{x}$ has a well defined frequency of oscillation, given by
$\omega_{1}\approx\epsilon_{0}$, with amplitude close to
$|\sigma_{x}(0)|$. Because of the weak coupling limit, the terms
in Eq. (\ref{sigmax}) due to the poles of the spectral density are
negligible compared to the others. Therefore, this explains why we
do not distinguish between several different regimes of
relaxation. The relaxation time $T_{1}$ is estimated to be of the
order of $1/\Gamma=12\mu s$. For the $\sigma_{z}$ component  we
see the expected beat structure, with well characterized beat and
angular frequencies given by $\omega_{1}$ and
$\Omega-\omega_{1}/2$, respectively.

The plots {\bf B} in figures \ref{sigxplot} and \ref{sigzplot}
present the dynamics for a case off resonance, $\Omega=0.8\Delta$.
Now we can see that several frequencies contribute to the spin
dynamics. For $\sigma_{x}$, the main oscillation frequency is
still $\omega_{1}$, however the amplitude of oscillations has
decreased more than one order of magnitude. The $\sigma_{z}$
dynamics has also changed. There is still a structure of beats but
more frequencies appear in the spectral decomposition.
$\omega_{1}$ and $\Omega+\omega_{1}/2$ are the beat and angular
frequencies for this case. For the very long time regime,
$t\gg1/\Gamma$, the component $\sigma_{x}$ reaches the stationary
equilibrium oscillating with frequency $2\Omega$ and very small
amplitude $G$, around the value $\sigma_{x}(\infty)$. For
$\sigma_{z}$ we observe that the beat structure disappears, giving
place to oscillations of frequency $\Omega$ and constant amplitude
$\frac{2\Delta}{\Omega}f(\epsilon_{0},\Omega)(\sigma_{x}(\infty)+G/2)$.
\\

In conclusion we have obtained approximate analytic solutions for
the Blcoh-Redfield equations of dissipative spins in a driving
field. The complex structure presented by the effective bath
spectral density brings a richer dynamical behavior for the spin
trapped in a quantum dot. We showed that the existence of the peak
in the spectral density introduces a new temporal scale in the
problem, $1/\ell$, the peak's width, which clearly separates
distinct time evolutions of the component $\sigma_{x}$. The first
one, short times, $t\ll1/\ell$, is basically determined by the
characteristics of the peak. We saw that the oscillations
presented in this regime have their natural frequencies shifted by
the bath resonance, $\Omega_s$, and their lifetimes are determined
by $1/\ell$. Thus, in principle, this regime could be used to
infer intrinsic quantities of the system-bath coupling , such as
the electron-phonon constant $\delta_s$. On the other hand, the
long time regime, $t\gg1/\ell$, was shown to be governed by the
external fields. We found that the relaxation rate $\Gamma$ for
this regime presents a strong peak for frequencies close to the
bath resonance, revealing the signature of the effective spectral
density. In addition, we were able to identify the shift induced
by the system-bath coupling  in the natural frequencies of the
system. We found that this shift, $\gamma$, has the form of a
temperature dependent Lamb shift \cite{mandel} \\

\section{Acknowledgments}

FB was supported by Funda\c{c}{\~a}o de Amparo {\`a} Pesquisa do
Estado de S{\~a}o Paulo (FAPESP), contract 01/05748-6. HW and AOC
acknowledge support from the Conselho Nacional de Desenvolvimento
Cient{\'\i}fico e Tecnol{\'o}gico(CNPq) and Hewlett-Packard
Brasil. AOC also thanks The Millenium Institute for Quantum Information. We 
would like to thank David DiVincenzo for a critical reading of the 
manuscript. 

\section{Appendix}

In this section we present the explicit form for the Bloch-Redfield coefficients used in the main text.

The first step is to construct the matrix elements of the non-dissipative time evolution operator $U(t,t')$,
\begin{widetext}
\begin{eqnarray}
\text{Im}U_{RR}(t,t')U_{RL}(t,t')&=&\frac{\cos^2(\phi)}{4}\bigg(\sin[(t-t')\Omega]+\sin[(t+t')\Omega]
(1-\cos[(t-t')\epsilon_1])\bigg)\nonumber\\
&&+\frac{(1+\sin(\phi))^2}{8}\sin[(t-t')(\Omega+\epsilon_1)]+\frac{(1-\sin(\phi))^2}{8}\sin[(t-t')
(\Omega-\epsilon_1)],\nonumber\\
\text{Re}U_{RR}(t,t')U_{RL}(t,t')&=&\frac{\cos(\phi)}{2}\sin(t'\Omega)\sin[(t-t')\epsilon_1]
-\frac{\sin(2\phi)}{4}\cos(t'\Omega)(1-\cos[(t-t')\epsilon_1]),\label{ele2}\\
|U_{RR}(t,t')|^2-|U_{RL}(t,t')|^2&=&\frac{\cos^2(\phi)}{2}\bigg(\cos[(t-t')\Omega]+\cos[(t+t')\Omega]
(1-\cos[(t-t')\epsilon_1])\bigg)\nonumber\\
&&+\frac{(1+\sin(\phi))^2}{4}\cos[(t-t')(\Omega+\epsilon_1)]-\frac{(1-\sin(\phi))^2}{4}
\cos[(t-t')(\Omega-\epsilon_1)].\nonumber
\end{eqnarray}\end{widetext}
If the limit $\frac{J(\omega)}{\omega}|_{\omega\rightarrow0}$
exists, and $J(\omega)$ does not have poles in the real axis, we
can write the real and imaginary parts of $M(t)$ as
${\rm{Re}}[M(t)]=\frac{1}{\pi}\int_0^\infty d\omega
J(\omega)\cos(\omega t)\coth(\beta\hbar\omega/2)$ and ${\rm{Im}}
[M(t)]=-\frac{1}{\pi}\int_0^\infty d\omega J(\omega)\sin(\omega
t)$, respectively. Now we are in position to calculate the
coefficients $A_i(t)$ and $\Gamma_{ij}(t)$. As pointed out
previously, if we first perform the time integrals, we can reach
very useful expressions for the frequency analysis of our problem.
Following this procedure, we obtain the fluctuation coefficients
$A_{x}$ and $A_{y}$ and the rate $\Gamma_{xx}(t)$ as (with
analogous expressions for $\Gamma_{xz}(t),$ and $\Gamma_{yz}(t)$)
\begin{widetext}
\begin{eqnarray}
A_x(t)&=&-\frac{\cos^2(\phi)}{2\pi}\bigg(1-\cos(2\Omega t)\bigg)I_1^{(0)}[\Omega;J(x),t]-
\left(\frac{\cos^2(\phi)\cos(2\Omega t)+(1+\sin(\phi))^2}{4\pi}\right)I_1^{(0)}
[\Omega+\epsilon_1;J(x),t]\nonumber\\
&&-\left(\frac{\cos^2(\phi)\cos(2\Omega t)+(1-\sin(\phi))^2}{4\pi}\right)I_1^{(0)}
[\Omega-\epsilon_1;J(x),t]-\frac{\cos^2(\phi)}{2\pi}\sin(2\Omega t)\bigg(I_2^{(0)}
[\Omega;J(x),t]\nonumber\\
&&-\frac{1}{2}(I_2^{(0)}[\Omega+\epsilon_1;J(x),t]+I_2^{(0)}
[\Omega-\epsilon_1;J(x),t])\bigg),
\end{eqnarray}
\begin{eqnarray}
A_y(t)&=&\frac{\sin(2\phi)}{2\pi}\bigg(\sin(\Omega t)I_1^{(0)}
[\Omega;J(x),t]+\cos(\Omega t)I_2^{(0)}[\Omega;J(x),t]\bigg)\nonumber\\
&&-\frac{\cos(\phi)}{2\pi}\sin(\Omega t)\bigg((1+\sin(\phi))I_1^{(0)}
[\Omega+\epsilon_1;J(x),t]-(1-\sin(\phi))I_1^{(0)}[\Omega-\epsilon_1;J(x),t]\bigg)\nonumber\\
&&-\frac{\cos(\phi)}{2\pi}\cos(\Omega t)\bigg((1+\sin(\phi))I_2^{(0)}
[\Omega+\epsilon_1;J(x),t]-(1-\sin(\phi))I_2^{(0)}[\Omega-\epsilon_1;J(x),t]\bigg),\\
\Gamma_{xx}(t)&=&\frac{\cos^2(\phi)}{2\pi}\left[\bigg(1+\cos(2\Omega t)\bigg)I_3
[\Omega,T,t]-\sin(2\Omega t)\bigg(I_4[\Omega,T,t]
-\frac{1}{2}(I_4[\Omega+\epsilon_1,T,t]+I_4[\Omega-\epsilon_1,T,t])\bigg)\right]\\
&&-\left(\frac{\cos^2(\phi)\cos(2\Omega t)-(1+\sin(\phi))^2}{4\pi}\right)I_3
[\Omega+\epsilon_1,T,t]-\left(\frac{\cos^2(\phi)\cos(2\Omega t)-
(1-\sin(\phi))^2}{4\pi}\right)I_3[\Omega-\epsilon_1,T,t],\nonumber
\end{eqnarray}
where we have defined the integrals
\begin{eqnarray}
I_1[y,t]&\equiv&\frac{1}{2}\int_0^\infty d\omega J(\omega)
\left(\frac{\sin((\omega-y)t)}{\omega-y}-\frac{\sin((\omega+y)t)}{\omega+y}\right),\nonumber\\
I_2[y,t]&\equiv&\int_0^\infty d\omega J(\omega)
\left(\frac{\sin^2((\omega-y)\frac{t}{2})}{\omega-y}+
\frac{\sin^2((\omega+y)\frac{t}{2})}{\omega+y}\right).\label{integrals1}\\
I_3[y,T,t]&\equiv&\frac{1}{2}\int_0^\infty d\omega
J(\omega)\coth\left(\frac{\hbar\omega}{2k_B T}
\right)\left(\frac{\sin((\omega-y)t)}{\omega-y}-
\frac{\sin((\omega+y)t)}{\omega+y}\right),\nonumber\\
I_4[y,T,t]&\equiv&\int_0^\infty d\omega
J(\omega)\coth\left(\frac{\hbar\omega}{2k_B T}\right)
\left(\frac{\sin^2((\omega-y)\frac{t}{2})}{\omega-y}+
\frac{\sin^2((\omega+y)\frac{t}{2})}{\omega+y}\right).\label{integrals2}
\end{eqnarray}
\end{widetext}

The special form of the integrals $I_{n}$ allows us to find
approximate analytic expressions, which are asymptotically exact
for $t\rightarrow\infty$. Considering that $J\left(\omega\right)$
satisfies the conditions
$J\left(0\right)=J\left(\infty\right)\rightarrow0$, we can write
for $t\gg1/\Omega_{c}$,
\begin{widetext}
\begin{eqnarray}
I_{1}\left(y,t\right)  \approx  \frac{\pi}{2}J\left(y\right)+
\frac{\pi}{2}\left\{ \text{Res}\left[J\left(z_{0}\right)\frac{e^{it(z_{0}-y)}}{z_{0}-y}\right]+\text{Res}
\left[J\left(z_{0}\right)\frac{e^{-it(z_{0}-y)}}{z_{0}-y}\right]\right\}\nonumber\\
 -\frac{\pi}{2}\left\{ \text{Res}\left[J\left(z_{0}\right)
 \frac{e^{it(z_{0}+y)}}{z_{0}+y}\right]+\text{Res}
 \left[J\left(z_{0}\right)\frac{e^{-it(z_{0}+y)}}{z_{0}+y}\right]\right\}, \label{i1}
\end{eqnarray}
\begin{eqnarray}
I_{2}\left(y,t\right)\approx\frac{1}{2}\int_{0}^{\Omega_{c}}d\omega
J\left(\omega\right)\text{P}\left[\frac{1}{\omega-y}+\frac{1}
{\omega+y}\right]-\frac{i\pi}{2}\left\{ \text{Res}
\left[J\left(z_{0}\right)\frac{e^{2it(z_{0}-y)}}{z_{0}-y}\right]-
\text{Res}\left[J\left(z_{0}\right)\frac{e^{-2it(z_{0}-y)}}{z_{0}-y}\right]\right\}\nonumber\\
-\frac{i\pi}{2}\left\{ \text{Res}\left[J\left(z_{0}\right)
\frac{e^{2it(z_{0}+y)}}{z_{0}+y}\right]-\text{Res}
\left[J\left(z_{0}\right)\frac{e^{-2it(z_{0}+y)}}{z_{0}+y}\right]\right\}, \label{i2} \end{eqnarray}
\end{widetext}
and similar expressions for $I_{3}$ and $I_{4}$. The residues in
the above expressions are calculated for the poles of $J\left(\omega\right)$
inside of the semi-circle centered at $z=\Omega_{c}$ and radius $\Omega_{c}$%
\footnote{The contour of integration should be chosen to ensure the convergence
of the integrals}. In this calculation, we have assumed that $y$ is not a pole
of the function $J\left(\omega\right)$. Here we have replaced the
upper limit of integration by the cutoff frequency $\Omega_{c}$.

Expressions (\ref{i1}) and (\ref{i2}) show us that the
time-dependent terms of $I_{n}$ have lifetimes determined by the
imaginary part of the poles of $J\left(\omega\right)$ in the
complex plane. In our case, the poles of $J\left(\omega\right)$
correspond to those of $J_{eff}\left(\omega\right)$ defined by
Eq.\ref{Jeff}.

\begin{figure*}
\begin{center}\includegraphics[%
  width=1.5\columnwidth]{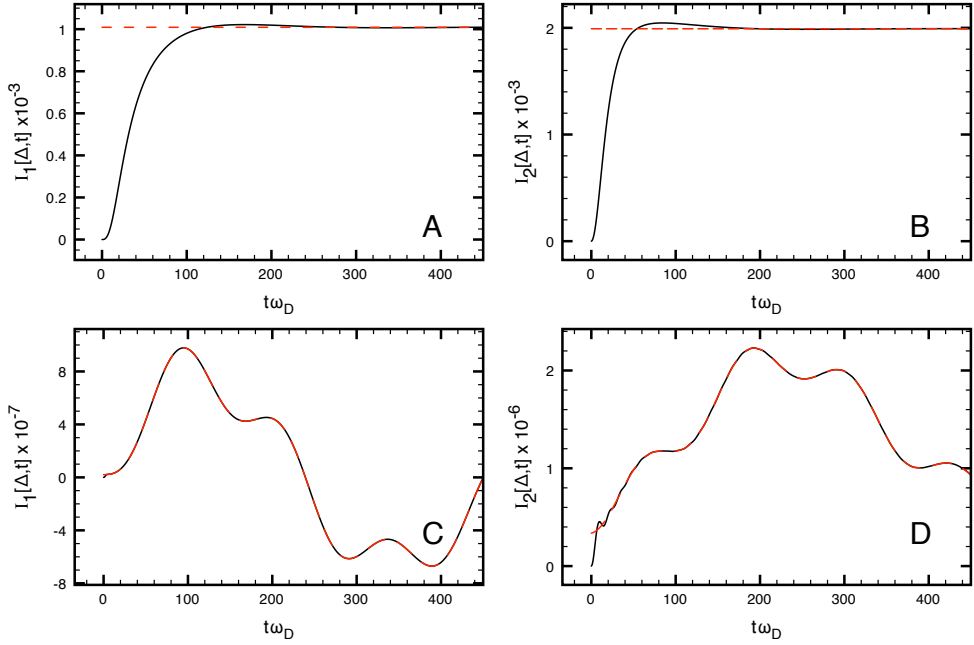}\end{center}
  \vspace{-0.6cm}
\caption{\label{i1graficos} $I_{1}^{(0)}[y;g(\omega),t]$ and
$I_{2}^{(0)}[y;g(\omega),t]$ integrals evaluated using the
analytic solutions Eqs.(\ref{i1}-\ref{i2}) (dashed lines) and the
exact numeric calculations (solid lines). Plots \textbf{A} and
\textbf{B} present the integrals computed for an Ohmic function
$g(\omega)=0.05\omega e^{-20\omega}$. Plots \textbf{C} and
\textbf{D} assume $g(\omega)$ as the spectral density
$J_{eff}(\omega)$ (Eq. \ref{Jeff}).}
\end{figure*}

Figure \ref{i1graficos} sketches the $I_1$ and $I_2$ behaviors
assuming $J(\omega)$ as an Ohmic function $J(\omega)=0.05\omega
e^{-20\omega}$ and the spectral density $J_{eff}(\omega)$ (Eq.
\ref{Jeff}). Comparing the results using the analytic form Eqs.
\ref{i1}-\ref{i2} and the exact numeric calculation, we find a
good agreement for $t\gg1/\omega_D$, specially for the case
$g(\omega)=J_{eff}(\omega)$ where the terms due the poles dominate
the short time regime, $t\leq1/\ell$.

\end{document}